\documentclass[aps,prb,reprint,onecolumn,a4paper,superscriptaddress,citeautoscript]{revtex4-1}
\usepackage{color}
\usepackage[english]{babel}
\usepackage[T1]{fontenc}
\usepackage{amsmath}
\usepackage{units}
\usepackage[charter]{mathdesign}
\usepackage[pdftex]{graphicx}
\begin{document}
\title{Electromagnetic properties of viscous charged fluids}
\author{Davide Forcella}
\affiliation{Physique Th\'eorique et Math\'ematique and International Solvay Institutes, Universit\'e Libre de Bruxelles, C.P. 231, 1050 Bruxelles, Belgium}
\author{Jan Zaanen}
\affiliation{The Instituut-Lorentz for Theroretical Physics, Leiden University, Leiden, The Netherlands}
\author{Davide Valentinis}
\author{Dirk van der Marel}
\affiliation{Department of Quantum Matter Physics, Universit{\'e} de Gen{\`e}ve, 24 quai Ernest-Ansermet, 1211 Gen{\`e}ve 4, Switzerland}
\begin{abstract}
We provide a general theoretical framework to describe the electromagnetic properties of viscous charged fluids, consisting for example of electrons in certain solids or plasmas.  We confirm that finite viscosity leads to multiple modes of evanescent electromagnetic waves at a given frequency, one of which is characterized by a negative index of refraction, as previously discussed in a simplified model by one of the authors. 
In particular we explain how optical spectroscopy can be used to probe the viscosity. We concentrate on the impact of this on the coefficients of refraction and reflection at the sample-vacuum interface. Analytical expressions are obtained relating the viscosity parameter to the reflection and transmission coefficients of light. We demonstrate that finite viscosity has the effect to decrease the reflectivity of a metallic surface, while the electromagnetic field penetrates more deeply. While on a phenomenological level there are similarities to the anomalous skin effect, the model presented here requires no particular assumptions regarding the corpuscular nature of the charge liquid. A striking consequence of the branching phenomenon into two degenerate modes is the occurrence in a half-infinite sample of oscillations of the electromagnetic field intensity as a function of distance from the interface.
\end{abstract}
\maketitle
%
%\tableofcontents
%
\onecolumngrid
\section{Introduction}
The flow properties of everyday fluids like water are governed by the Navier-Stokes theory of hydrodynamics. The key parameter governing the 
dissipative aspects of such fluids are the bulk- and shear viscosities. The most abundant electrically charged fluids are formed by electrons in metals.
The theory of transport in {\em normal} metals is well understood. These form Fermi-liquids; although at precisely zero temperature this supports a
"collisionless" quantum hydrodynamics which is noticeably different from the classical Navier Stokes hydrodynamics at any finite temperature and at sufficiently 
long times the gas of thermally excited quasiparticles takes over forming yet again a "collision-full" effective classical fluid \cite{landau1959}. 
These notions were successfully verified in the 1960's and 1970's in the neutral Fermi liquid realised in $^3$He, in the form of the famous maximum
in the attenuation of zero sound \cite{roach1976}. The zero sound of the collisionless regime corresponds with a coherent vibration of the Fermi surface with 
a damping proportional to the microscopic collision rate of the quasiparticles $1 /\tau_{coll} \sim T^2$, while in the classical regime the damping is 
just viscous. The damping is in turn determined by 
$\Gamma \simeq (E_F/m_e) \tau_{coll}$, where $E_F$ is the 
Fermi energy \cite{forster1995}. 

The shear viscosity of the Fermi liquid is given by $\eta \simeq n E_F \tau_{coll}$ and  it is noticed that the viscosity (and attenuation) are now proportional to the quasiparticle collision time, diverging like $1/T^2$ at
low temperatures. Why is it so that this quantity does not play any role in the transport theory of electrons in  metals? The reason is well understood. For a hydrodynamical description to make sense the conservation of total momentum of
the fluid flow in a finite density system is required at least on the time and length scales associated with the establishment with local equilibrium. 
This in turn requires Galilean invariance at least on the microscopic scale and in normal metals this is explicitly broken by the atomic lattice. Even 
when this is perfectly periodic, it is detrimental for the momentum conservation in a Fermi liquid. The reason is that the quasiparticles are characterised  by a large Fermi-momentum $k_F$ which is of order of the Umklapp wavevector $\vec{Q}$ with the effect that the states at $k_F$ have always a finite admixture of Umklapp copies. This in turn has the effect that at every microscopic collision a certain amount of momentum is dumped in the lattice  expressed in terms of the "Umklapp efficiency" $\Delta$, such that the (microscopic) momentum relaxation rate  $1/\tau_{K} = \Delta /\tau_{coll}$  while $\Delta\sim 0.5$ in the transition metals\cite{lawrence1973}. 

Momentum conservation is therefore already destroyed at the microscopic cut-off in normal metals. More concretely this implies that in an expansion of the time and coordinate derivatives of the current density with $\tau_K $ as the expansion parameter, the leading term gives the dissipation of the local current density proportional to $1/\tau_K$, followed by terms proportional to dissipation caused by gradients of the current density. The latter terms thus represent a correction on the leading momentum dissipation, where the constant of proportionality defines $\eta$. One of the challenges yet to be met is to determine the value of $\eta$ in a non-Gallilean invariant setting, and its dependence on temperature and frequency. In the context of this paper we will simply assume that this higher order correction exists, and work out a number of physical consequences. For this we will adopt values for $\eta$ which seem plausible at this moment, but which need to be determined ultimately on the basis of first principles and/or experimental data.  
As discussed recently by Andreev, Kivelson and Spivak it might well be that  an exception is presented in electron systems with a very low density, where $k_F$ becomes very small, and density of scattering centers is small compared to $1/Q$. In this case other interesting transport properties could emerge due to interaction of the scattering centers with the viscous charged fluid\cite{andreev2011}.    

Very recently  it has been questioned to what extend these notions apply to {\em non-Fermi liquids} which might be realised in the form  of strange metals as in high T$_c$ superconductors, heavy fermion systems and so forth. It is believed that these are governed by quantum  criticality. At zero temperature their quantum dynamics would be scale invariant while the expectation is that at finite temperatures these are characterised by extremely short "Planckian" \cite{zaanen2004} relaxation times $\tau_{\hbar} \sim \hbar / {k_B T}$ \cite{sachdev1999,muller2009}.   
This got further impetus by the discovery of the "minimal viscosity-entropy" ratio using the AdS/CFT correspondence\cite{policastro2001} which seems confirmed both in the quark gluon plasma created at the heavy ion colliders\cite{rebhan2012} as well as the cold atom unitary Fermi gas\cite{schafer2009}. 
It is possible that such quantum critical systems are much less sensitive to Umklapp scattering since these lack intrinsic microscopic scales like $1/k_F$. It was argued that it could therefore well be that {\em first} hydrodynamics is realised in such 
systems while only at longer time scales the lack of Galilean invariance becomes noticeable \cite{hartnoll2012}.
Resting on these ideas, several proposals appeared recently for the explanation of the mysterious linear resistivity of for instance the normal state
of the high T$_c$ superconductors \cite{davison2014,hartnoll2014a,zeng2014}. 

Electron systems in metals can be probed most conveniently by electromagnetic fields.  The question arises, what would be the fingerprint of (non) hydrodynamical behavior of the electrons in the electromagnetic responses that can be measured in the laboratory? Perhaps surprisingly, to the best of our knowledge a general theoretical framework to describe the electromagnetic effects of finite viscosity for charged fluids is not available. The goal of this paper is to construct such a general phenomenological description.  
In the context of heavy fermion superconductivity, sound attenuation experiments have also been discussed as a probe of the viscosity tensor\cite{pethick1986,moreno1995}. The existence of at least two different methods for probing the viscosity would allow to calibrate the various different methods against each other.
In Refs. \onlinecite{amariti2011} and \onlinecite{amariti2013} the effect of viscosity on the refractive index was discussed for different systems, using a string theory setup in the former and linear response theory for pure hydrodynamics plus electromagnetism in the latter. In particular a prediction was given for the generic presence of negative refraction that should be manifest in actual systems such as a quark-gluon plasma. In  Ref. \onlinecite{ALW3} the existence of multiple electromagnetic waves with the same frequency was addressed. 
In Ref. \onlinecite{Forcella:2014dwa} these works were extended for intrinsically strongly coupled materials beyond the hydrodynamic limit, using the AdS/CFT correspondence.
In principle these phenomena could also be present in the electron liquid for certain parameter ranges. For the interpretation of practical experiments two more steps are needed: generalize the theoretical framework to include retarded effect for the viscosity and the presence of a lattice or impurities; relate transmission or reflection coefficients to the parameters describing the electron liquid. The purpose of the present paper is to fill this gap and describe the phenomenology of charged fluids.

The basic strategy of the approach in this paper is to provide a general equation describing the viscous dynamics of the transverse velocity of the fluid and couple it to Maxwell equations to arrive to a theory of the electromagnetic response. Such  a closed system leads {\em inter alia} to negative refraction and a bifurcation into two modes inside the viscous charge liquid instead of one at any given frequency. In the paper we apply this formalism to the response of a Fermi liquid with finite momentum relaxation to a transverse force\cite{roach1976}, where it is easy to incorporate the viscous response on a phenomenological level, however it should be understood that our theoretical framework and its main consequences apply more broadly, without the need to stack on a specific model. The description of transmission and reflection at the interface involves three constituant relations at the interface instead of the usual two relations. The corresponding modification of the Fresnel equations is derived in analytical closed form. The resulting reflection coefficient exhibits a peak for $\omega\approx 1/\tau_K$, which disappears in the limit of zero viscosity. The transmission through a thin metallic film exhibits an increased transparency if the electronic viscosity is finite, which is accompagnied by a strong frequency dependence of the phase which is a sensitive parameter of the viscosity. Finally, in the limit of a weakly interacting Fermi liquid the surface impedance corresponds closely to the results for the anomalous skin effect obtained in the Reuter-Sondheimer approach\cite{reuter1948,sondheimer2001}, obtained with the same parameters. The model presented here can be regarded as a generalization of the Reuter-Sondheimer model to viscous charge liquids at large. 
\section{Equations of Transverse Motion of a Viscous Charged Fluid }
\subsection{Electromagnetic field coupled to a viscous charged fluid}\label{MNS}
The velocity field, $\vec{\upsilon}(t,\vec r)$, of a liquid of particles of charge $e$, mass $m$ and density $n$ gives rise to an electrical current density $\vec J(t,\vec r) =  n e  \vec{\upsilon}(t,\vec r)$. Here we will concentrate on fields polarized transverse to the direction of propagation corresponding to the wave vector $\vec q$. The wave propagation of the transverse electric ($E^T$)  fields is described by the Maxwell equations. 
For the velocity distribution of the currents (${\upsilon}^T$) \cite{footnote1} we need to take into account the inertial response to a force, the force exerted by the electric field, the momentum relaxation rate to the crystal lattice, and the non-local coupling within the fluid. The non-local corrections can be written as an expansion in the coordinate derivatives of the velocity field. The leading non-local correction for the transverse velocity field is then given by the Laplacian $\Delta  {\upsilon}(t, \vec r)$. If a given flow pattern is suddenly switched off at $t=0$, a non-equilibrium situation will persist on the time-scale of the inter particle collision time $\tau_{coll}$. Consequently the individual particles respond to the flow pattern that has existed in the past, {\em i.e.} the force, is $\int_{-\infty}^0 M_{\nu}(t) \Delta  {\upsilon}(t, \vec r)dt$, where $M_{\nu}(t)$ is the memory function related to the viscosity.  In the frequency domain this can be represented as $\nu(\omega)\Delta  {\upsilon}(\vec r)$, where, according to the preceding argument, the generalized viscosity $\nu(\omega)$ is a causal response function corresponding to the Fourier transform of $M_{\nu}(t)$. For fields and currents with a time dependence described by $\exp{(-i\omega t)}$ the corresponding set of differential equations forms a simple closed system
\begin{eqnarray}
\label{M}
\left[c^2\Delta +{\omega^2}\right]E^{}(\vec r) &=&   -4 \pi i \omega 
J(\vec r) \\
\label{NS}
\left[\tau_{K}^{-1} -i  \omega  - \nu(\omega) \Delta  \right]{\upsilon}^{} &=&  e E^{}(\vec r)/m 
\end{eqnarray}
For $\tau_{K}^{-1}=0$ the second equation is the Navier-Stokes equation for transverse currents, generalized to a frequency dependent $\nu(\omega)$. 
The eigenstates of Eqs. \ref{M} and \ref{NS} are linear superpositions of terms of the form $\exp{(i{q}_{j}z-i\omega t)}$ having the same frequency $\omega$, and $q_{j}$ satisfies the self-consistent relation 
\begin{equation}\label{disprel}
\frac{q^2 c^2}{\omega^2}
= 1- \frac{\omega_p^2}{ \omega   (\omega  +i  \tau_{K}^{-1} + i {\nu(\omega)} q^2) } 
\end{equation}
where we defined $\omega_p^2\equiv {4\pi n e^2}/{m}$. 
The ${q}^2$ dependence of the pole results from the non-zero viscosity of the medium and gives rise to all the exotic effects that we describe in this paper. For $\nu(\omega)=0$ Eq. \ref{disprel}
reduces to the dielectric function described by the Drude model. 
Previously Benthem and Kronig have derived a similar relation, where they neglected the displacement current as being small compared with the conduction current (Eq. (6) of Ref. \onlinecite{benthem1954}). From this expression they calculated the surface impedance, assuming for $\nu$ the universal value $\sim \hbar / m$.
Combining Maxwell's equations with the general expression for the convective derivative for the velocity of an electron, Gilberd arrived at a different result \cite{gilberd1982}, where in the denominator of the expression for $q^2 c^2 /\omega^2$ a real term { proportional to $q/\omega$ appeared instead of the imaginary dispersive term $i {\nu} q^2$} of Eq. \ref{disprel} \cite{notegilberd1982}. 
A similar situation exists when light is absorbed by excitations exhibiting a non-negligible dispersion as a function of $q$. This is known to occur for excitons in semiconductors \cite{cocoletzi2005}, and has been predicted for strongly dispersing optical phonons \cite{helm2003}. Unlike Eq. \ref{disprel},  in these cases the $q$-dispersion enters through a non-dissipative term in the  { denominator} of the dielectric function. 
Using the relation $q_{j}=n_{j}k$ between wave vector and refractive index where $k\equiv\omega/c$ { is the wavenumber in vacuum}, and solving Eq. \ref{disprel} for ${q}_{j}$,  we obtain 
\begin{equation}\label{disprel2}
2n_j^2
=
1-\frac{1-i\omega\tau_{K}}{\omega^2\nu_c\tau_{K}}
\pm 
\sqrt{\left[1+\frac{1-i\omega\tau_{K}}{\omega^2\nu_c\tau_{K}}\right]^2
+\frac{i4\omega_p^2}{\omega^3 \nu_c}}
\end{equation}
where we adopted the compact notation $\nu_c=\nu(\omega)/c^2$ and $n_j$ and $\nu_c$ have implicit frequency dependence. The electro-hydrodynamical properties are thus characterized by the two time scales $\nu_c$ and $\tau_{K}$, and by the plasma frequency $\omega_p$. The first interesting observation is that for any given frequency $\omega$, {\em two} modes of electromagnetic field coupled to matter (labeled $j=1,2$) exist in a viscous charged liquid, which are distinguished by the two possible values of $n_j^2(\omega)$, a phenomenon described as additional light waves in Ref. \onlinecite{pekar1958}. 
\subsection{Behaviour at the vacuum/matter interface}\label{interfaces}
Here we concentrate on experiments which can be performed with state of the art methods, namely reflection at the surface of a sample and transmission through a film. 
The Maxwell equations provide the conditions that at each interface $E(z)$ and $\partial E/\partial z$  are continuous. An additional condition follows from the properties of Newtonian fluids. The tangential friction per unit area exerted by the moving fluid on the boundary of the solid is, in leading order, proportional to the velocity at the interface, $\kappa {\upsilon}$. In equilibrium this has to be balanced by the force exerted by the velocity gradient of the viscous fluid, $\eta \partial {\upsilon}/\partial z$, leading to the Navier { constitutive relation}\cite{bocqueta2007}. Taken together we arrive at the following three { constitutive relations} at the two interfaces
\begin{align}% left aligned
\label{boundary}
E(0-\delta)&=E(0+\delta)&E(d-\delta)&=E(d+\delta)
\nonumber \\
\left.\frac{\partial E}{\partial z}\right|_{0-\delta}&=\left.\frac{\partial E}{\partial z}\right|_{0+\delta}
&\left.\frac{\partial E}{\partial z}\right|_{d-\delta}&=\left.\frac{\partial E}{\partial z}\right|_{d+\delta}
\nonumber \\
\frac{1}{\lambda}&=\left.\frac{\partial \ln \upsilon}{\partial z}\right|_{0+\delta}
&
\left.\frac{\partial \ln \upsilon}{\partial z}\right|_{d-\delta}&=\frac{-1}{\lambda}
\end{align}
The constant  $\lambda=\eta / \kappa$ is the slip length, where $\lambda=0$  ($\lambda=\infty$) corresponds to the interface being maximally rough (smooth). 
To be specific we consider an electromagnetic wave of frequency $\omega$ propagating along $z$ from $-\infty$ to the sample, which has one boundary defined by the plane $z=0$ and the other by $z=d$. Part of the wave is reflected back, with an amplitude characterized by the reflection coefficient $r$, the amplitude transmitted to $z>d$ is characterized by the transmission coefficient $t$, and inside the slab the wave-amplitude is a superposition of the 4 modes:
\begin{align}\label{rtt}
&E(z)/E(0)&=&
 e^{ik z}+re^{-ik z}&(z<0) \nonumber \\
&&=&t_1 e^{in_1 k z}+\theta_1 e^{-in_1 k z}+t_2 e^{in_2 k z}+\theta_2 e^{-in_2 k z}  &(0<z<d)\nonumber \\
&&=&te^{ik z} &(z>d)
\end{align}
{ Since} Im$n_j$>0, the two terms $e^{-in_j k z}$ are exponentially diverging for $z\rightarrow\infty$. In the limit of a half infinite sample $\theta_1$ and $\theta_2$ therefore converge to zero, and only $t_1$ and $t_2$ contribute to the transport of electromagnetic radiation into the material. 

In the case of reflection/transmission at the vacuum/sample interface of a half-infinite sample we combine  Eqs. \ref{rtt} with aforementioned { constitutive relations} at the vacuum-matter interface at $z=0$, which leads in a straightforward fashion to the transmission and reflection coefficients at such an interface
\begin{eqnarray}\label{transmission,reflection}
t_{j}&=&
\frac{2(n_{\underline{j}}-1)}{(n_{{j}}+1)(n_{\underline{j}}-n_{j})}
\frac{1-n_{\underline{j}}i\lambda k}
{1+(1-n_{\underline{j}}-n_{j})i\lambda k}\nonumber
\nonumber \\
r&=&t_1+t_2-1 .
\end{eqnarray}
The surface impedance $Z$ is obtained from the second member of Eq.  \ref{transmission,reflection}, using the general expression relating surface impedance and reflection coefficient
\begin{equation}\label{Z}
\frac{Z}{Z_0}=\frac{1+r}{1-r}
\end{equation}
where $Z_0$ is the vacuum impedance. 

Another relevant case is that of a film of thickness $d$ with vacuum on either side. The field inside the film is a superposition of all 4 solutions of Eq. \ref{disprel2}, {\em i.e.} the exponentially decaying as well as the exponentially increasing ones. Taken together with the reflection amplitude $r_{film}$  for $z<0$ and the transmission amplitude $t_{film}$ for $z>d$ the problem of the constitutive relations at both interfaces corresponds to a system of 6 linear equations with 6 unknown parameters. By combining the constitutive relations defined in Eq. \ref{boundary}, four of these combinations provide the matrix expression
\footnotesize{
\begin{equation}\label{eq:matrix1}
  {\left[ \begin{matrix}
  1+n_1 & 1-n_1 & 1+n_2 & 1-n_2 \\
  (1-n_1)f_1 & (1+n_1)/f_1 & (1-n_2)f_2& (1+n_2)/f_2 \\
  (1-n_1^2)(1-n_1  \xi) & (1-n_1^2)(1+n_1  \xi) & (1-n_2^2)(1-n_2  \xi) & (1-n_2^2)(1+n_2 \xi)  \\
  (1-n_1^2)(1+n_1  \xi)f_1 & (1-n_1^2)(1-n_1  \xi)/f_1 & (1-n_2^2)(1+n_2  \xi)f_2 & (1-n_2^2)(1-n_2 \xi)/f_2  \\
  \end{matrix}
  \right]}
 \left[ \begin{matrix}
  t_1 \\
  \theta_1 \\
  t_2 \\
  \theta_2 \\
  \end{matrix}
  \right]  
  =
  \left[ \begin{matrix}
  2 \\
  0 \\
  0 \\
  0 \\
  \end{matrix}
  \right]
%\nonumber
  \end{equation}  
}\normalsize
where we use the compact notations  $f_1=e^{i n_1 k d}$, $f_2=e^{i n_2 k d}$ and $\xi=i\lambda k$.  Numerical inversion provides $t_1$, $\theta_1$, $t_2$ and $\theta_2$, from which the reflection and transmission coefficients of the film are obtained using the remaining two constitutive relations
\begin{equation}\label{eq:t}
\begin{aligned}
&r_{film}= t_1  + \theta_1  + t_2  + \theta_2 -1\\
&\frac{t_{film}}{t_{vac}}=e^{-i k d}\left\{ t_1 e^{i n_1 k d} + \theta_1 e^{-i n_1 k d} +t_2 e^{i n_2 k d} + \theta_2 e^{-i n_2 k d}\right\}
\end{aligned}
\end{equation}
where, similar as in experimental practice, the transmission is calibrated against the transmission through a slice of vacuum with the same thickness, $d$, as the film. 
\section{Relevant parameter range of momentum relaxation and viscosity coefficient. }
Before we turn to the examples based on numerical solution of the expressions in the previous section, it will be useful to explore the relevant parameter range for the viscosity and the relaxation rate. Two limiting cases have been explored theoretically in some detail in the literature: Fermi-liquids and quantum critical states. 

Since in a Fermi-liquid context viscosity and diffusivity correspond to the same quantity\cite{forster1995}, it therefore follows that $\nu(0)=\upsilon_F^2 \tau_{coll}$. For our numerical examples we make the reasonable  approximation that the memory function follows an exponential decay, {\em i.e.}  $M_{\nu}(t)\tilde{\tau}=\nu(0) \exp{(-t/\tilde{\tau})}$, so that $\nu(\omega)=\nu(0) / (1-i\omega\tilde{\tau})$. Definitions of $\tilde{\tau}$, $\tilde{\upsilon}_F$ and further details are provided in the Appendix. When we solve Eq. \ref{NS} with this function, 
we obtain the relations for transverse sound in a Fermi liquid describing the dispersion (Re $q(\omega)$) and the attenuation (Im $q(\omega)$) of first ($\omega\tau<<1$) and zero ($\omega\tau>>1$) transverse sound of liquid $^3$He in the normal state.  We thus obtain the following expression for the generalized viscosity 
\begin{equation}\label{mem_viscosity}
\frac{\nu_c(\omega)}{\tilde{\tau}}\sim \frac{(\tilde{\upsilon}_F/c)^2}{1-i\omega \tilde{\tau}}
\end{equation}
The result for $\nu(\omega)$ in the hydrodynamic limit $(\omega\rightarrow 0)$ is a real number, hence the viscous response is purely dissipative. In the "collisionless" limit, {\em i.e.} for frequencies high compared to the collision rate, this crosses over to purely reactive response.  Moreover $\nu(\omega)$ is  proportional to $\upsilon_F^2$, implying that materials with a high Fermi velocity such as aluminum (having $\upsilon_F/c\sim 0.003$) are expected to be record holders for viscosity related phenomena. 

Despite its equally high Fermi velocity a drastically different situation has been anticipated for graphene as a result of the quantum criticality of this system\cite{muller2009}, therefore bringing it close to the lower bound conjectured in the context of the quark gluon plasma.
In this case the dynamic viscosity (related to the kinematic viscosity as $\eta=mn\nu$)
is given by the relation \cite{kovtun2005} $ \eta / s \ge A \hbar / k_B$ where $s$ is the entropy density. We associate an elastic mean free path $l_0$ with the breaking of the Galilean
invariance, and the kinematic viscosity $\nu$ with the diffusivity. 
Besides the assumption that intrinsic rapid relaxation processes are governing the electron system, an additional condition is that the length scales associated with the momentum relaxation processes are still large. Operationally  this means that $l_0$ has to be large compared to the lattice constant: this is a  "clean limit" notion. The transport of quantum critical systems can also be addressed -to a degree- in the dirty limit \cite{hartnoll2014b} where very different principles are at work.
The momentum relaxation rate is in such simple liquids (and also the local quantum critical liquid computed holographically in Ref. \onlinecite{davison2014}) 
$1/\tau_{K} = \nu/l_0^2$. Taken together these arguments than lead to the simple result
\begin{equation}\label{nuQCmuFL}
\frac{\nu_c}{\tau_{K}  } \sim \left(\frac{\lambda_e}{2 l_0}\right)^2 \left(\frac{s}{nk_B}\right)^2
\end{equation}
where $\lambda_e=2.4 \cdot 10^{-14}$ cm is the Compton wavelength for electrons. Since $l_0$ must be larger than the interatomic distance, $l_0 > 10^{-8} $ cm and $s$ can not exceed the equipartition value, $s < k_B n $, we conclude that $\nu_c/\tau_{K} < 10^{-12}$, {\em i.e.} some 7 orders of magnitude below the Fermi liquid estimate. It may therefor be difficult in practice to experimentally access the regime relevant to the quantum critical state. For a Fermi liquid on the other hand, the parameters are much more favorable for experimental observation.
\section{Numerical examples}
\begin{figure}[]
\includegraphics[width=\columnwidth]{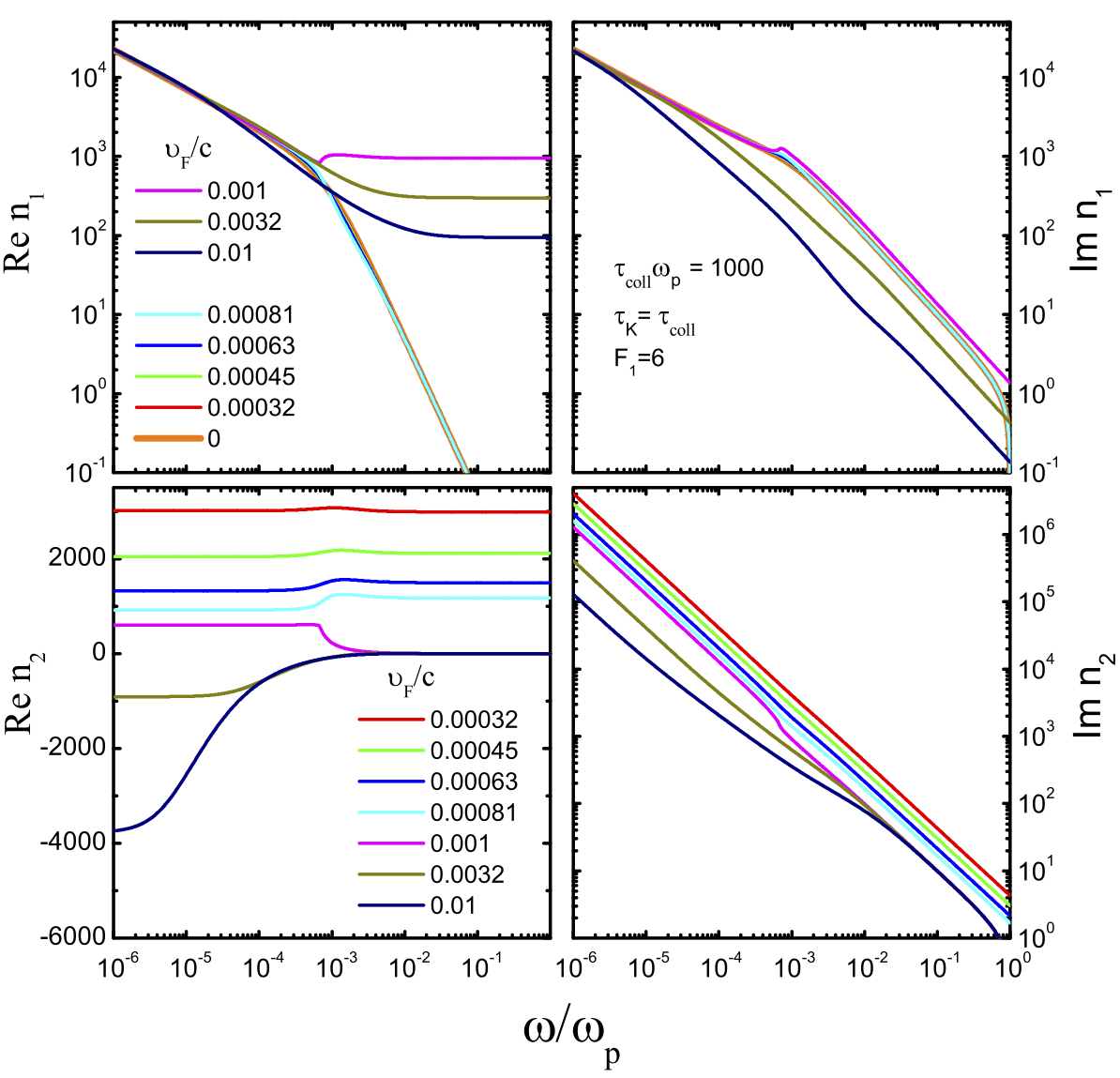}
\caption{\label{fig1}The real and imaginary part of the refractive indices $n_1$ and $n_2$, as a function of the normalized frequency for selected values of the Fermi velocity in units of light velocity, $\upsilon_F/c$. In all panels the relaxation time $\tau_{K}=1000 /\omega_p$. In the two top panels showing Re$n_1$ and Im$n_1$ the curves for $\upsilon_F/c= 0.00032$ coincide within plotting accuracy with the Drude result, $\upsilon_F=0$.}
\end{figure}
\subsection{Refractive indices}
In Fig. \ref{fig1} the real and imaginary part of the two refractive indices are displayed for selected values of the parameters describing the fluid, utilising Eq. \ref{viscosity2} for the frequency dependence of the viscosity parameter. For the range $\omega\tau_{K}<1$ the two solutions of Eq. \ref{disprel} are described by the leading order terms of the expansion in $\nu_c\omega$ and $\omega\tau_{K}$
\begin{eqnarray}
n_1&\approx&\omega_p\sqrt{\frac{i\tau_{K} }{\omega}}\left(
1-\frac{i}{2}\nu_c\omega\tau_{K}^2\omega_p^2\right)
\nonumber\\
n_2&\approx&\sqrt{\frac{\tau_{K}}{4\nu_c}}\left( 1-\nu_c\tau_{K}\omega_p^2\right)
+\frac{i}{\omega\sqrt{\nu_c\tau_{K}}}
\end{eqnarray}
In the non-viscous limit $n_1$ converges to the conventional expression of an evanescent electromagnetic wave: In this limit the second mode has Im$n_2\rightarrow \infty$, so that the wave amplitude, $\exp{(-\mbox{Im} n_2 k z)}$, vanishes for all $z$. Consequently the $n_1$ mode is the sole non-trivial solution for $\nu=0$. For the viscous case there exist two solutions for the same $\omega$, corresponding to collective modes of different admixtures of the coupled charge-liquid and the electromagnetic field. At low frequencies the $n_1$ branch approaches the usual evanescent electromagnetic wave. The second branch $n_2$ has, by virtue of the constant real part, the characteristics of transverse sound, with an attenuation constant diverging as $1/\omega$. This mode is similar in character to aforementioned attenuated transverse sound of $^3He$\cite{roach1976}. Another intriguing aspect of this mode is the fact that for $\nu_c \tau_{K} \omega_p^2 > 1$, the real part of $n_2$ is negative at low frequencies. This is the footprint of the very interesting phenomenon called negative refraction \cite{veselago1968,agranovich2006,pendry2000}, in which the phase velocities and the energy flux are in opposite direction. 
It was discussed in the context of charged fluids in Refs. \onlinecite{amariti2011} and \onlinecite{amariti2013}, where a particular form of equations \ref{M},\ref{NS} was considered, for which the viscosity is real and frequency independent, and the momentum is not dissipated: $1/ \tau_{K}=0$.
The frequency $\omega\tau_{K}=1$ constitutes a peculiar bifurcation point: for $\omega\tau_{K}>1$ the character of $n_1$ and $n_2$ is swapped when the viscosity drops below the critical value $\nu_c \tau_{K} \omega_p^2=1$; {\em i.e.} for $\nu_c \tau_{K} \omega_p^2<1$ the $n_1$ branch has Re$n_1\sim 1$ and  Re$n_2 >> 1$, whereas for $\nu_c \tau_{K} \omega_p^2>1$ this is the other way around.
\subsection{Surface impedance}
\begin{figure}[]
\includegraphics[width=\columnwidth]{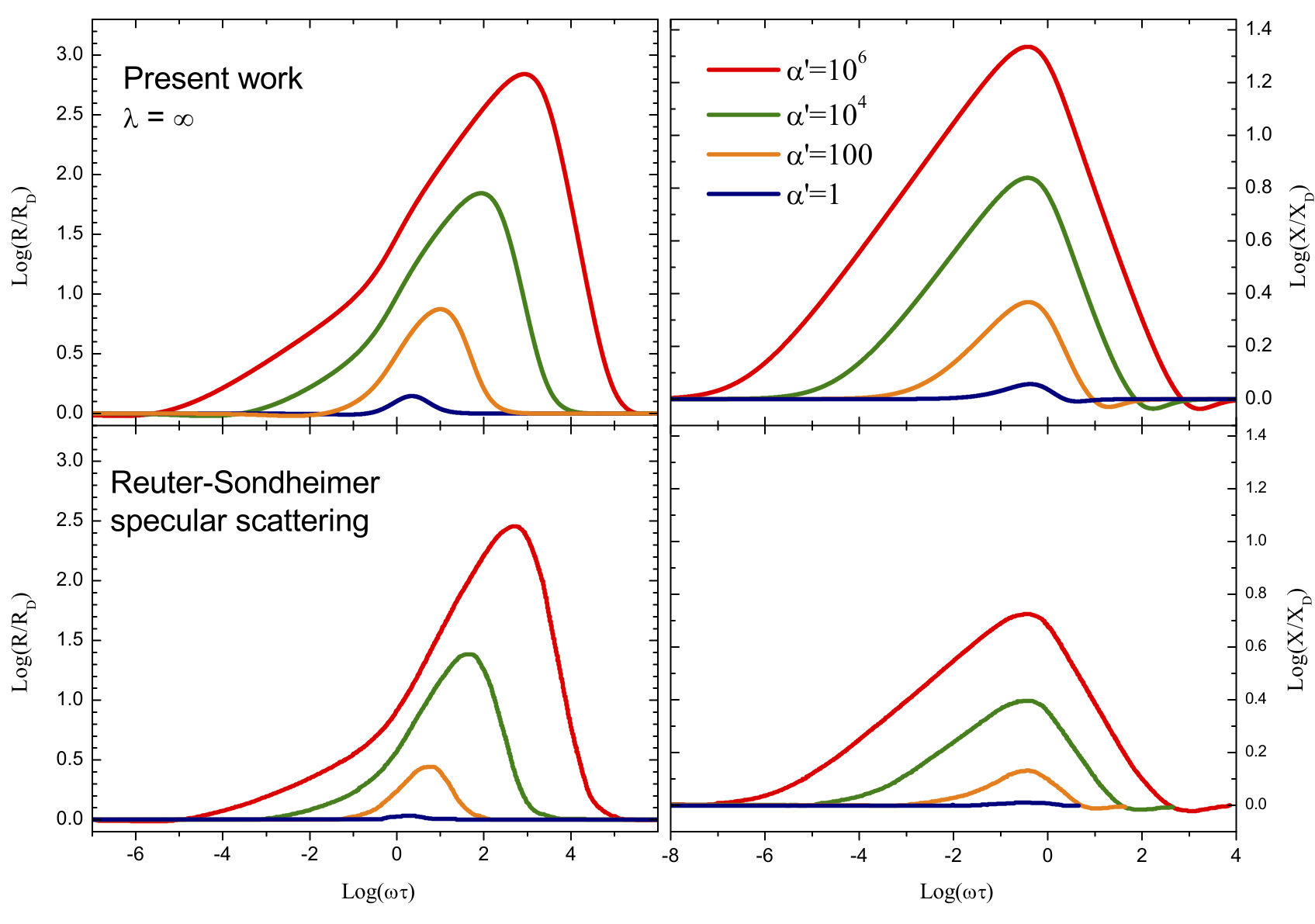}
\caption{\label{fig2} Frequency dependence of the surface resistance (left) and reactance (right) relative to the Drude limit  ($\upsilon_F=0$) for different values of the parameter $\alpha^{\prime}=(3/4)( \tau_{coll}\omega_p\upsilon_F/c)^2$. The calculations were done for $\tilde{\tau}=\tau_K=\tau_{coll}$ and a frictionless surface ($\lambda=\infty$). For comparison the results for the limit of specular scattering of Reuter and Sondheimer (Figs. 2 and 3 of Ref.\onlinecite{reuter1948}) are reproduced in the lower panel on the same scale and for the same values of $\alpha^{\prime}$.}
\end{figure}
In Fig. \ref{fig2} the resulting spectra of $R/R_D$ and $X/X_D$ represent the real and imaginary parts of the surface impedance relative to the corresponding Drude limit  ($\upsilon_F=0$). Ordinates and abscissas are shown on a $^{10}$log scale. The results in Fig. \ref{fig2} are in fact almost identical to the calculations of Reuter and Sondheimer, displayed in Fig. 3 of Ref. \onlinecite{reuter1948} for the same parameter choices of $\alpha^{\prime}=(3/4)(\tau\omega_p\upsilon_F/c)^2$. As is the case in Ref. \onlinecite{reuter1948} the resulting curves for any given value of $\alpha^{\prime}$ are universal, {\em i.e.} they do not depend on the particular choice of $\upsilon_F/c$ and $\tau\omega_p$. The Reuter-Sondheimer model parts from a weakly interacting electron m¥odel, for which the Boltzmann equations are solved in the case where the mean free path is longer than penetration of the electromagnetic rays. The situation at the surface is treated in terms of a fraction $p$ of particles which are scattered specularly, and $1-p$ which is scattered diffusively. The results in Fig. 3 Ref. \onlinecite{reuter1948} are for the limit of pure specular scattering, corresponding to a perfectly smooth interface for which $\lambda=\infty$. The close correspondence between the Reuter-Sondheimer prediction and the present result implies that, for the case of weakly interacting electrons, we have obtained an alternative formulation of the anomalous skin effect, with a set of simple expressions in analytically closed form. At the same time, the present approach has a potentially broader applicability since it does not rely on any particular assumptions regarding the corpuscular nature of the charge liquid, and is of particular interest for the optical properties of  quantum critical matter.
\subsection{Reflection at vacuum/matter interface of a half-infinite sample}
\begin{figure}[]
\includegraphics[width=\columnwidth]{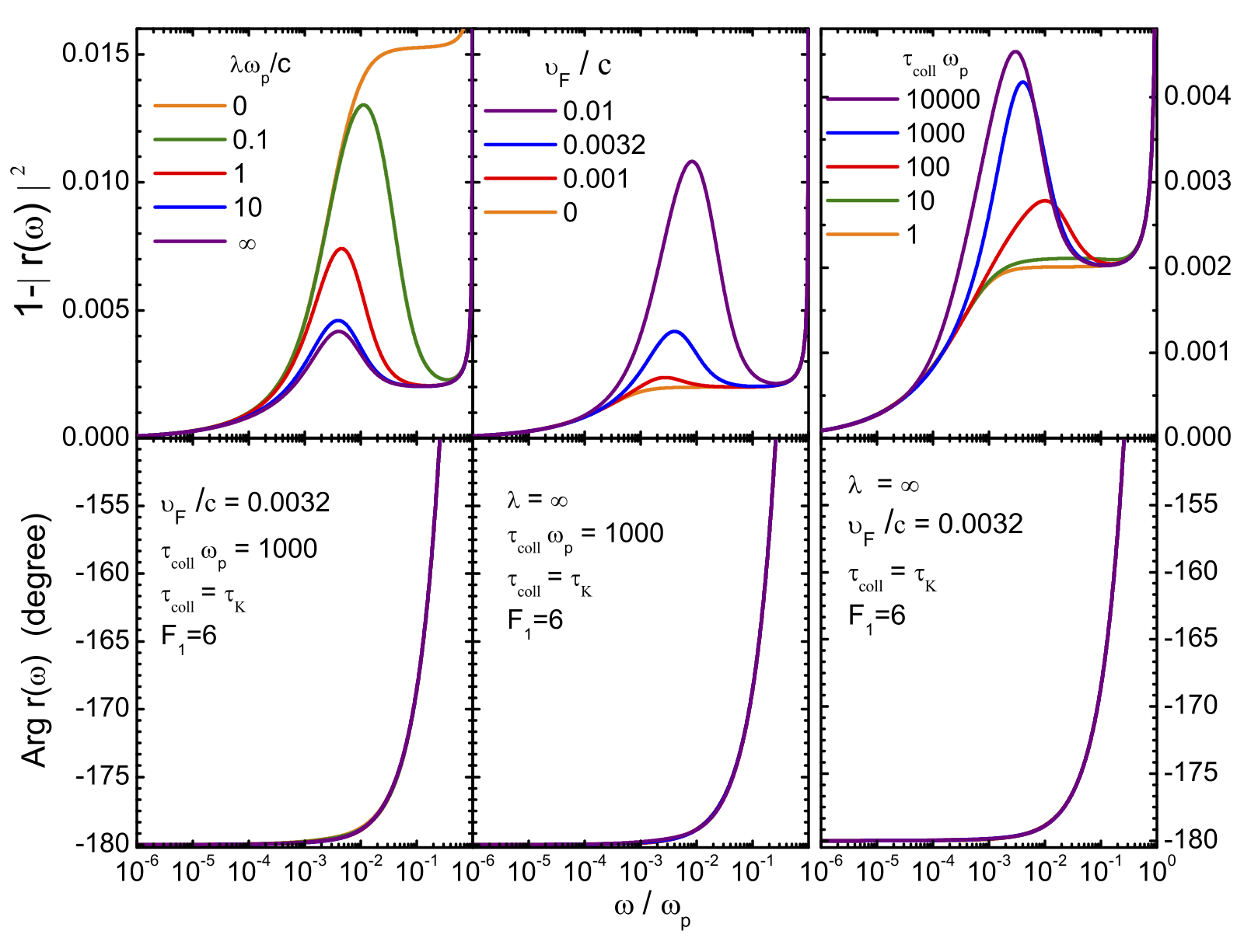}
\caption{\label{fig3} Spectra of the absorption coefficient ($A=1-|r|^2$) 
%on a logarithmic scale, 
and the phase (in degrees) of the reflection coefficient as a function of frequency for selected values of the slip length ${\lambda}$ (left), relative Fermi velocity $\upsilon_F/c$ (middle), and momentum relaxation time, $\tau_{K}\omega_p$ (right).}
\end{figure}
The result for the reflection coefficient is displayed in Fig. \ref{fig3} for selected values of the parameters. For a clean material with a perfectly smooth surface, the expected behavior corresponds to the result shown in the rightmost panel with $\lambda\omega_p=\infty$. Based on the estimates made above for the viscosity and relaxation time, the curve with $\upsilon_F/c= 0.0032$ comes closest to the expected parameter range of a Fermi liquid. We should contrast this curve to the Drude curve (orange). Clearly, the viscosity has the effect of suppressing the reflection coefficient, or increasing the absorption on the solid: the absorption forms a peak at $\omega\tau_{K}\approx 1$, where the maximum increases from about 0.002 in the Drude limit to 0.004 for $\upsilon_F/c= 0.0032$. Increasing the surface roughness (left panel of Fig. \ref{fig3}) demonstrates the increased effects of finite viscosity on the absorption of the material. Intuitively this confirms what one expects: Extreme surface roughness forces the current to be zero at the sample surface. Due to the viscosity this slows also the current deeper in the fluid, which therefor is less effective in screening the external electromagnetic field, and so becomes a less effective mirror.  The phase shown in the lower panels is a much less sensitive probe of the viscosity parameter. Since the phase of a reflectivity signal is already notoriously difficult to measure, for all practical purposes the best strategy for experiments is probably to concentrate on the absorption coefficient $1-|r(\omega)|^2$. 
\subsection{Non-exponential decay and amplitude oscillations inside the viscous charge liquid.}
\begin{figure}[]
\includegraphics[width=\columnwidth]{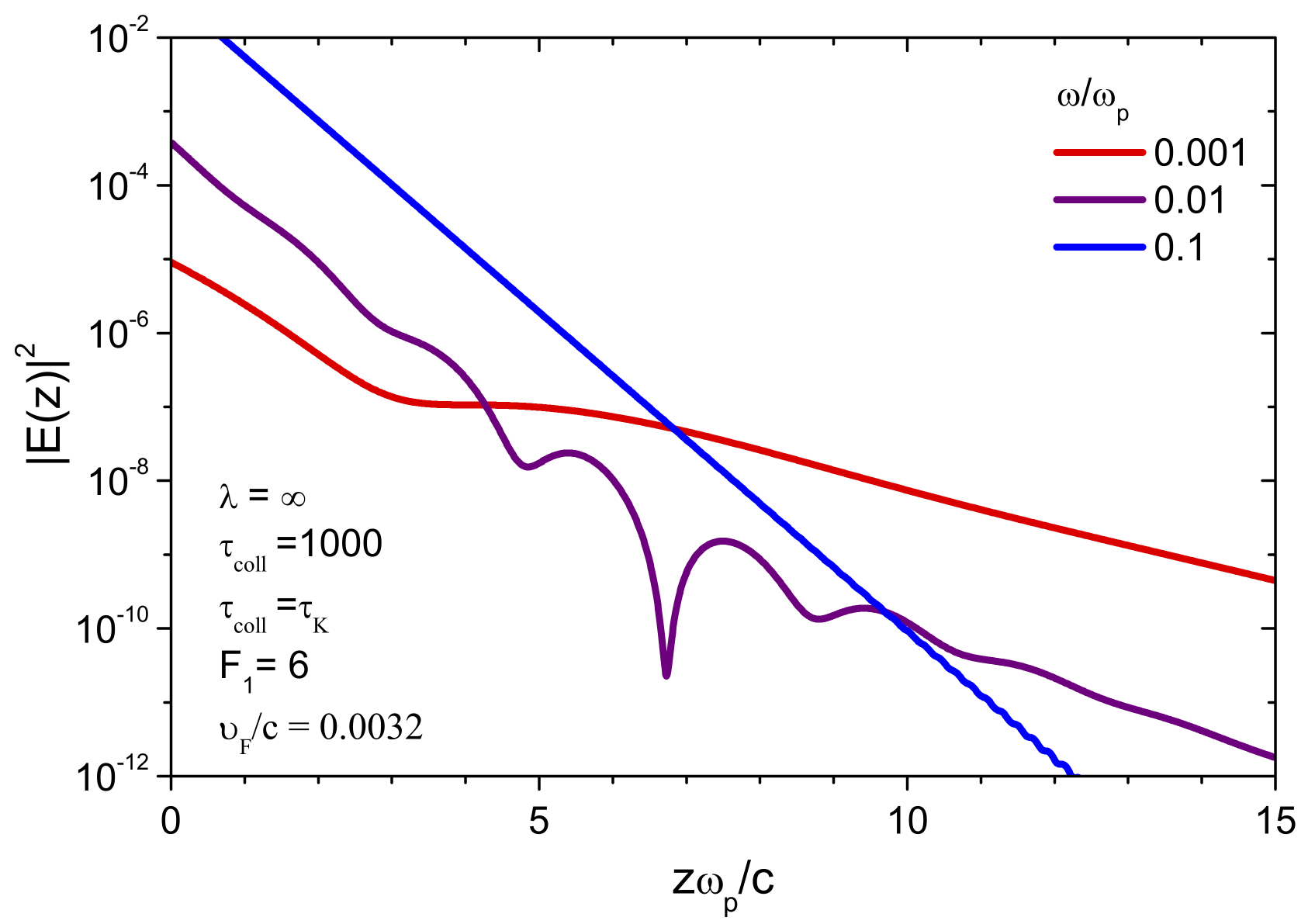}
\caption{\label{fig4} Oscillations of the electric field intensity resulting from interference by the two degenerate modes of the same frequency, as a function of penetration away from the vacuum/sample interface. }
\end{figure}
An electromagnetic wave of frequency $\omega$ incident at the surface will excite two modes of the same frequency inside the material, with amplitudes given by the transmission coefficients $t_j$ at the vacuum-matter interface. Both solutions are exponentially decaying as a function of distance, each with a different decay length. An additional consequence is that, since $n_1$ and $n_2$ have different real parts, the intensity $|t_1\exp{(i n_1  kz)} + t_2\exp{(i n_2  kz)}|^2$ exhibits standing wave patterns as a function of distance from the interface in the range where both modes are of comparable amplitude. An example of this behavior is shown in Fig. \ref{fig4}. In principle experimental methods can be devised to measure these oscillations of the field by local probe techniques.  
\subsection{Thin film transmission}
\begin{figure}[]
\includegraphics[width=\columnwidth]{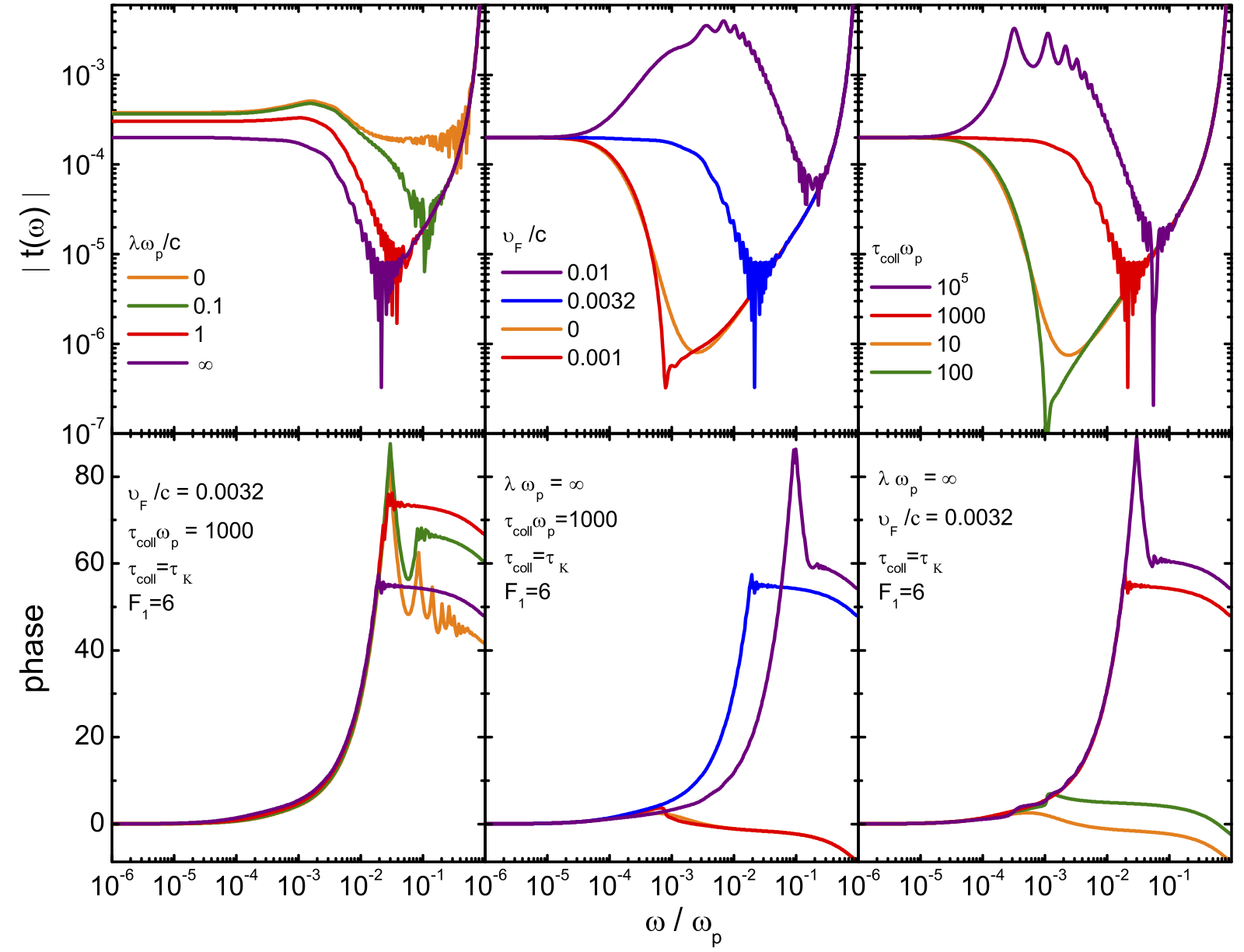}
\caption{\label{fig5} Spectra of amplitude and phase (in radians) of the transmission of a film of viscous charged fluid of thickness $d=10c/\omega_p$ for selected values of the slip length ${\lambda}$ (left), relative Fermi velocity $\upsilon_F/c$ (middle), and collision time, $\tau_{coll}$ (right). The phase values are calibrated against vacuum, {\em i.e.} if the sample is replaced with vacuum the phase is zero at all frequencies. }
\end{figure}
As pointed out at the end of section \ref{interfaces}, the thin film transmission can be calculated numerically from the reflection and transmission amplitudes at both interfaces using Eq. \ref{eq:t}, and a numerical inversion of the $4\times 4$ matrix of Eq. \ref{eq:matrix1} describing the 4 rays inside the film. Examples for representative parameters are shown in Fig. \ref{fig5}. Just as for the reflectivity of a thick sample, the viscosity has the effect of changing the spectral appearance: for the Drude model without viscosity the transmission spectrum has a plateau for $\omega\tau_{K}<1$, followed by a minimum at $\omega\tau_{K}\approx 1$ and a rise to about $|t|=1$ for $\omega \rightarrow \omega_p$. Increasing the viscosity has several effects: The transmission develops a maximum at $\omega\tau_{K}\approx 1$, the minimum is pushed to somewhat higher frequencies, and also in this case the transmission rises to $|t|=1$ for $\omega \rightarrow \omega_p$. Interestingly, in the frequency range between the maximum and the minimum there are Fabry-Perot resonances, associated with the $n_1$ branch which has a relatively weak attenuation in this frequency range. Note that the propagation of the $n_1$ mode becomes similar to a standard polariton mode in an insulating material despite that we are dealing with a metal.
We also see that the phase of the transmission spectrum is an extremely sensitive probe of the viscosity for $\omega\tau_{K} >> 1$. In particular, while for small values of $\nu_c(0)$ it is a weakly decreasing function of frequency, it has a rapid rise for $\upsilon_F/c>0.003$. 
\section{Conclusions}
We have derived a general framework to deal with electromagnetic properties of charged fluids and we have derived some of their main properties. We have shown the existence of two coupled electromagnetic-matter modes for a given frequency  inside a viscous charged fluid, one of these two has a negative value for the real part of the index of refraction for low frequency. Moreover we have computed some observables: reflection and transmission coefficients, that could be measured in actual experiments.
Our results provide perspectives for a novel generation of experiments and devices for the detection of viscosity in charged or electron liquids and the possible exploitation of multiple waves and negative refraction in such systems.
\section{Acknowledgements}
DvdM acknowledges illuminating discussions with Jean-Louis Barrat, Teun Klapwijk and Pierce Coleman. This work was supported by the Swiss National Science Foundation (SNSF) through Grant 200021-153405. JZ acknowledges support of a grant from the John Templeton foundation. The opinions expressed in this publication are those of the authors and do not necessarily reflect the views of the John Templeton Foundation. 
DF is F.R.S-FNRS Charg\'e de Recherches. He acknowledges support by the F.R.S.-FNRS, by IISN - Belgium through conventions 4.4511.06 and 4.4514.08, by the Communaut\'e Francaise de Belgique through the ARC program and by the ERC through the SyDuGraM Advanced Grant. DF acknowledges the kind hospitality of the LPTHE, where part of this research has been implemented.
\appendix
\section{Neutral Fermi liquid\label{appendix:A}}
For our numerical examples we adopt the dispersion $q(\omega)$ of transverse sound of the neutral Fermi liquid, which is obtained in two steps\cite{abrikosov1959,roach1976}. First $x$ is solved for a given (real) frequency $\omega$ from the transcendental relation
\begin{equation}\label{FL_viscosity2}
(1-x^2) \left(1-\frac{x}{2}\ln\left[\frac{x-1}{x+1}\right]  \right)
= \frac{1+(1-i\omega\tau_{coll})(F_1-6)/9}{1+(1-i\omega\tau_{coll})F_1/3}
\end{equation}
where in general the solutions for $x$ have a complex value. Here $F_1$ is the Landau parameter characterizing the interaction in the  $l=1$ angular momentum channel, and Landau parameters of the higher angular momentum channels are assumed to be negligible. In the second step $q(\omega)$ is calculated from 
\begin{equation}\label{FL_viscosity2b}
q(\omega)=\frac{i+\omega\tau_{coll}}{x\upsilon_F\tau_{coll}}.
\end{equation}
We see, that the function $q(\omega)$ depends uniquely on the collision rate $\tau_{coll}^{-1}$, the Fermi-velocity $\upsilon_F$, and on $F_1$. 
We obtained the following parametrization by fitting to the numerical solution of Eq. \ref{FL_viscosity2}: 
\begin{equation}\label{dispersion}
q^2\upsilon_F^2\tau_{coll}^2=5i\omega\tau{_{coll}}(1-i\omega\tau_{coll} \sqrt{F_1}7/32)/(1+F_1/3)
\end{equation} 
This expression merges with the exact solution for $\omega\rightarrow 0$, and it is rather accurate for all other frequencies and parameters.

To obtain the viscosity from this, we observe that the hydrodynamic free propagation of transverse polarized modes in a neutral liquid is described by 
$\partial \upsilon / \partial t =  \nu\Delta \upsilon$, from which $i\omega=\nu q^2$. In order to extend the description to finite frequencies beyond the hydrodynamic limit, we introduce the frequency dependent memory function 
\begin{equation}\label{viscosity1}
\nu(\omega)=\frac{i\omega}{q^2}
\end{equation}
with $q(\omega)$ given by either the solution of Eq. \ref{FL_viscosity2} or the {parametrization, Eq. \ref{dispersion}.} In the latter case we obtain the expression
\begin{equation}\label{viscosity2}
\nu(\omega)=\frac{\upsilon_F^2\tau_{coll} (1+F_1/3)/5}{1-i\omega\tau_{coll} \sqrt{F_1}7/32}
\end{equation}
We recognize here Eq. \ref{mem_viscosity}, where the various parameters are related as
\begin{align}
&\tilde{\tau}=\frac{7\sqrt{F_1}}{32}\tau_{coll}\nonumber\\
&\tilde{\upsilon}_F^2=\upsilon_F^2 \frac{1+F_1/3}{\sqrt{F_1}}\frac{32}{35}\nonumber\\
&\frac{1}{{\tau}_K}=\frac{7\sqrt{F_1}}{32} \frac{\Delta}{\tilde{\tau}}
\end{align}

\begin{thebibliography}{99}
\bibitem{footnote1}The remainder of the discussion is entirely devoted to transverse polarized fields. For the sake of compactness of notation we will drop the suffix $T$ from here on.
\end{thebibliography}
\end{document}